\pdfoutput=1
\documentclass[sigconf,authorversion=true,nonacm]{acmart}
\renewcommand\footnotetextcopyrightpermission[1]{} %
\acmConference[CCS '22]{CCS '22: 29th ACM Conference on Computer and Communications Security}{November 7-11, 2022}{Los Angeles, U.S.A.}
\acmYear{2022}

\usepackage{tikz}
\usepackage{pgfplots}
\usepackage[american]{babel}
\usepackage{amsmath}
\usepackage{xspace}
\usepackage{filecontents}
\usepackage[textsize=scriptsize]{todonotes}
\usepackage{pifont}
\usepackage{hyperref}
\usepackage{booktabs}
\usepackage{tikz}
\usepackage{tikzsymbols}
\usepackage{tikzpeople}
\usepackage{url}

\usepackage{enumitem}
\setlist{nosep}
\usepackage{subcaption}
\usepackage{booktabs}

\newcommand{\name}{GuaranTEE\xspace}
\newcommand{\target}{target\xspace}
\newcommand{\rogueadmin}{malicious administrator\xspace}
\newcommand{\rogueclient}{malicious client\xspace}

\newcommand{\analyzer}{analyzer\xspace}
\newcommand{\rlog}{attestation log\xspace}
\newcommand{\prover}{ProveTEE\xspace}
\newcommand{\verifier}{VerifyTEE\xspace}
\newcommand{\trampoline}{trampoline\xspace}
\newcommand{\queue}{queue\xspace}
\newcommand{\Queue}{Queue\xspace}
\newcommand{\callid}{ID\xspace}
\newcommand{\callids}{IDs\xspace}
\newcommand{\pthread}{prover thread\xspace}
\newcommand{\vthread}{verifier thread\xspace}
\newcommand{\batch}{ID batch\xspace}
\newcommand{\batches}{ID batches\xspace}
\newcommand{\Batches}{ID Batches\xspace}

\newcommand{\ff}{feedback frequency\xspace}
\newcommand{\bsize}{\batch size\xspace}
\newcommand{\sprovider}{service owner\xspace}
\newcommand{\sproviders}{service owners\xspace}

\usepackage[toc,acronym,nowarn]{glossaries}

\newacronym{SE}{SE}{Secure Element}
\newacronym{NFC}{NFC}{Near Field Communication}
\newacronym{BLE}{BLE}{Bluetooth Low Energy}
\newacronym{RIL}{RIL}{Radio Interface Layer}
\newacronym{SELinux}{SELinux}{Security-Enhanced Linux}
\newacronym{LSM}{LSM}{Linux Security Modules}
\newacronym{TCB}{TCB}{Trusted Computing Base}
\newacronym{MAC}{MAC}{Mandatory Access Control}
\newacronym{IPC}{IPC}{Inter-Process Communication}
\newacronym{ICC}{ICC}{Inter-Container Communication}
\newacronym{ISA}{ISA}{Instruction Set Architecture}
\newacronym{cgroups}{cgroups}{control groups}
\newacronym{CA}{CA}{Certificate Authority}
\newacronym{PKI}{PKI}{Public Key Infrastructure}
\newacronym{MITM}{MITM}{Man-In-The-Middle}
\newacronym{BYOD}{BYOD}{Bring-Your-Own-Device}
\newacronym{CM}{CM}{Container Management}
\newacronym{SM}{SM}{Security Management}
\newacronym{HAL}{HAL}{Hardware Abstraction Layer}
\newacronym{TLS}{TLS}{Transport Layer Security}
\newacronym{C2C}{C2C}{Container To Container}
\newacronym{Protobuf}{Protobuf}{Protocol Buffers}
\newacronym{SDO}{SDO}{Sensitive Data Object}
\newacronym{VMA}{VMA}{Virtual Memory Area}
\newacronym{PGD}{PGD}{Page Global Directory}
\newacronym{PUD}{PUD}{Page Upper Directory}
\newacronym{PMD}{PMD}{Page Middle Directory}
\newacronym[firstplural=Page Table Entries (PTEs)]{PTE}{PTE}{Page Table Entry}
\newacronym{COW}{COW}{Copy-On-Write}
\newacronym{IV}{IV}{Initialization Vector}
\newacronym{ESSIV}{ESSIV}{Encrypted Salt-Sector Initialization Vector}
\newacronym{KSM}{KSM}{Kernel Samepage Merging}
\newacronym{JIT}{JIT}{Just-In-Time}
\newacronym{DMA}{DMA}{Direct Memory Access}
\newacronym{FDE}{FDE}{Full Disk Encryption}
\newacronym{AS}{AS}{Address Space}
\newacronym{GCM}{GCM}{Google Cloud Messaging}
\newacronym{TPM}{TPM}{Trusted Platform Module}
\newacronym{JTAG}{JTAG}{Joint Test Action Group}
\newacronym{LUKS}{LUKS}{Linux Unified Key Setup}
\newacronym{VPN}{VPN}{Virtual Private Network}
\newacronym{PBKDF2}{PBKDF2}{Password-Based Key Derivation Function 2}
\newacronym{KVM}{KVM}{Kernel-based Virtual Machine}
\newacronym{HV}{HV}{Hypervisor}
\newacronym{SEV}{SEV}{Secure Encrypted Virtualization}
\newacronym{SME}{SME}{Secure Memory Encryption}
\newacronym{TSME}{TSME}{Transparent \gls{SME}}
\newacronym{SP}{AMD-SP}{Secure Processor}
\newacronym[firstplural=Guest Virtual Adresses (GVAs)]{GVA}{GVA}{Guest Virtual Address}
\newacronym[firstplural=Guest Physical Addresses (GPAs)]{GPA}{GPA}{Guest Physical Address}
\newacronym[firstplural=Host Physical Addresses (HPAs)]{HPA}{HPA}{Host Physical Address}
\newacronym[firstplural=System Physical Addresses (SPAs)]{SPA}{SPA}{System Physical Address}
\newacronym{GPT}{GPT}{Guest Page Table}
\newacronym{HPT}{HPT}{Host Page Table}
\newacronym{TLB}{TLB}{Translation Lookaside Buffer}
\newacronym{ORAM}{ORAM}{Oblivious RAM}
\newacronym{SEV-ES}{SEV-ES}{SEV Encrypted State}
\newacronym{SEV-SNP}{SEV-SNP}{SEV Secure Nested Paging}
\newacronym{RMP}{RMP}{Reverse Map Table}
\newacronym{VMCB}{VMCB}{Virtual Machine Control Block}
\newacronym{SLAT}{SLAT}{Second Level Address Translation}
\newacronym{SSH}{SSH}{Secure Shell}
\newacronym{RSA}{RSA}{Rivest–Shamir–Adleman}
\newacronym{ECDHE}{ECDHE}{Elliptic-Curve Diffie-Hellman Ephemeral}
\newacronym{AES}{AES}{Advanced Encryption Standard}
\newacronym{OOM}{OOM}{Out Of Memory}
\newacronym{MKTME}{MKTME}{Multi-Key Total Memory Encryption}
\newacronym{VMI}{VMI}{Virtual Machine Introspection}
\newacronym{MAD}{MAD}{Median Absolute Deviation}
\newacronym{HSM}{HSM}{Hardware Security Module}
\newacronym{AE}{AE}{Automatic Exit}
\newacronym{NAE}{NAE}{Non-Automatic Exit}
\newacronym{AES-NI}{AES-NI}{AES New Instructions}
\newacronym{NIC}{NIC}{Network Interface Card}
\newacronym{NMI}{NMI}{Non-Maskable Interrupt}
\newacronym{MTU}{MTU}{Maximum Transmission Unit}
\newacronym{VA}{VA}{Virtual Address}
\newacronym{GFN}{GFN}{Guest Frame Number}
\newacronym{SFN}{SFN}{System Frame Number}
\newacronym{IOMMU}{IOMMU}{I/O Memory Management Unit}
\newacronym{AISE}{AISE}{Address Independent Seed Encryption}
\newacronym{MT}{MT}{Merkle Tree}
\newacronym{BMT}{BMT}{Bonsai Merkle Tree}
\newacronym{LPID}{LPID}{Located Page IDentifier}
\newacronym{CB}{CB}{Counter Block}
\newacronym{PRD}{PRD}{Page Root Directory}
\newacronym{SWIOTLB}{SWIOTLB}{Software I/O Translation Buffer}
\newacronym{ASID}{ASID}{Address Space Identifier}
\newacronym{vCPU}{vCPU}{virtual CPU}
\newacronym{VC}{\texttt{\#VC}}{VMM Communication Exception}
\newacronym{GHCB}{GHCB}{Guest Hypervisor Communication Block}
\newacronym{IDT}{IDT}{Interrupt Descriptor Table}
\newacronym{ASLR}{ASLR}{Address Space Layout Randomization}
\newacronym{KASLR}{KASLR}{Kernel Address Space Layout Randomization}
\newacronym{SLES}{SLES}{SUSE Linux Enterprise Server}
\newacronym{RHEL}{RHEL}{RedHat Enterprise Linux}
\newacronym{IBS}{IBS}{Instruction Based Sampling}
\newacronym{CFM}{CFM}{Control Flow Modification}
\newacronym{CE}{CE}{Code Execution}
\newacronym{TSC}{TSC}{Time Stamp Counter}
\newacronym{TEE}{TEE}{Trusted Execution Environment}
\newacronym{CFA}{CFA}{Control-Flow Attestation}
\newacronym{CFI}{CFI}{Control-Flow Integrity}
\newacronym{SGX}{SGX}{Software Guard Extensions}
\newacronym{TA}{TA}{Target Application}
\newacronym{TZ}{TZ}{TrustZone}
\newacronym{HA}{HA}{Host Application}
\newacronym{BIOS}{BIOS}{Basic Input Output System}
\newacronym{ROP}{ROP}{Return-Oriented Programming}
\newacronym{TSX}{TSX}{Transactional Synchronization Extensions}
\newacronym{CFG}{CFG}{Control-Flow Graph}
\newacronym{DOP}{DOP}{Data-Oriented Programming}
\newacronym{DEP}{DEP}{Data Execution Prevention}
\newacronym{KDF}{KDF}{Key Derivation Function}

\makeglossaries
\glsdisablehyper

\addto\extrasamerican{

}
 
\graphicspath{{figures/}}

\newcommand{\phase}[1]{\ding{\numexpr181 + #1}}

\begin{document}

\date{}

\title{\name: Introducing Control-Flow Attestation for Trusted Execution Environments}

\author{Mathias Morbitzer}
\affiliation{
	\institution{Fraunhofer AISEC}
	\city{Garching near Munich}
	\country{Germany}
}
\email{mathias.morbitzer@aisec.fraunhofer.de}
\author{Benedikt Kopf}
\affiliation{
	\institution{Fraunhofer AISEC}
	\city{Garching near Munich}
	\country{Germany}
}
\email{benedikt.kopf@aisec.fraunhofer.de}
\author{Philipp Zieris}
\affiliation{
	\institution{Fraunhofer AISEC}
	\city{Garching near Munich}
	\country{Germany}
}
\email{philipp.zieris@aisec.fraunhofer.de}

\begin{abstract}
 The majority of cloud providers offers users the possibility to deploy \glspl{TEE} 
 to protect their data and processes from high privileged adversaries. 
 This offer is intended to address concerns of users when moving critical tasks into the cloud. 
 However, \glspl{TEE} only allow to attest the integrity of the environment at launch-time. 
 To also enable the attestation of a \gls{TEE}'s integrity at run-time, we present \name. 
 \name uses control-flow attestation to ensure the integrity of a service running within a \gls{TEE}. 
 By additionally placing all components of \name in \glspl{TEE}, 
 we are able to not only detect a compromised target, but 
 are also able to 
 protect ourselves from malicious administrators.
 We show the practicability of \name by providing a detailed performance and security evaluation of our prototype based on Intel SGX in Microsoft Azure. 
 Our evaluation shows that the need to transfer information between \glspl{TEE} and the additional verification process add considerable overhead under high CPU load. 
 Yet, we are able to reduce this overhead by securely caching collected information and by performing the analysis in parallel to executing the application. 
 In summary, our results show that \name provides a practical solution for cloud users focused on protecting the integrity of their data and processes at run-time.  
\end{abstract}

\maketitle

\section{Introduction}
\glsreset{TEE}

Cloud computing allows users to quickly adapt their IT infrastructures to today's ever-changing requirements. 
However, relocating infrastructure into the cloud 
means that users are required 
to trust the cloud provider, something especially businesses often find difficult~\cite{amigorena2019why}.
To cover these concerns, various \glspl{TEE} have been proposed by academia~\cite{suh2003aegis, evtyushkin2014isox, costan2016sanctum, weiser2019timber, lee2020keystone} and the industry~\cite{arm2009trustzone, boivie2013secureblue, costan2016intel,kaplan2016amd, intel2020tdx, hunt2021confidential, arm2021cca}. 
A \gls{TEE} aims to protect data and processes inside the \gls{TEE} from adversaries, even if they are located at higher privilege layers.
To ensure their trustworthiness, \glspl{TEE} provide mechanisms to attest that they were set up correctly and not manipulated before launch. 
Yet, such static attestation mechanisms only assess the \gls{TEE}'s state at launch-time and are not able to detect attacks during run-time. 
In other words, an attacker exploiting a vulnerability in the software running within the \gls{TEE}
will not be detected by static attestation. 

Cloud environments create an additional threat by requiring users to perform all operations
remotely. 
This dependence inherits the still widespread threat of remote code execution attacks~\cite{miller2019trends, ozkan2020vulnerability}. 
Remote code execution attacks are performed during run-time and aim to divert the execution flow of the target 
to perform malicious operations. 
To divert the target's execution flow, attackers most commonly resort to \emph{control flow attacks}, which overwrite code pointers in memory, such as a return address on the stack.
This, for example allows to perform \gls{ROP}~\cite{shacham2007geometry}, in which the attacker executes short sequences of instructions, the so-called gadgets. 
Carefully chaining those gadgets allows to perform Turing-complete computations. 
To 
detect such attacks, Abera et al. proposed \gls{CFA}~\cite{abera2016cflat}. 
\gls{CFA} records the control flow of a program and afterwards compares it to a set of previously determined legal control flows. 
However, most previous work on \gls{CFA} targets embedded systems, focusing on specific challenges for such environments, for example limited resources.   
In comparison, cloud environments have access to a vast amount of resources, while at the same time presenting different challenges, such as a contrasting usage model. 
Specifically, while embedded systems are mainly used by their owners, cloud applications are often designed to provide a service to autonomous clients. 
As those clients are often considered untrusted, performing \gls{CFA} in cloud environments requires moving the attestation away from the client to other, trusted entities. 

It is exactly this gap, the lack of \gls{CFA} designs adapted to cloud environments, which we aim to close with this work. 
To be precise, we present \name, a design that combines the security guarantees of \glspl{TEE} with run-time verification via \gls{CFA}, allowing us to detect control flow attacks within \glspl{TEE}. 
With these capabilities, we identify security critical microservices in a cloud environment serving requests to autonomous clients as one of many possible use cases.
By using \name, we are able to protect the microservices running in the untrusted cloud environment from a high privileged adversary in control of the underlying infrastructure. 
Additionally, we are able to detect clients exploiting software vulnerabilities at run-time. 
To demonstrate the practicability of \name, we describe our prototype, which we implemented using Intel SGX and evaluated in Microsoft Azure.
We will open source our prototype and our LLVM extensions once the paper is set to be published.  

In summary, we make the following contributions: 
\begin{itemize}
  \item We present \name, a design for cloud environments which adds \gls{CFA} to \glspl{TEE}, allowing us to detect attacks that modify the \gls{TEE}'s control flow.
   \item We demonstrate the practicability of our design by presenting a prototype based on the widespread \gls{TEE} Intel SGX. 
  \item We describe our extensions to the LLVM compiler framework which allow for the easy deployment of Intel SGX enclaves protected with \name.
  \item We provide a detailed performance evaluation of the \name prototype in Microsoft Azure using the \texttt{sgx-nbench} benchmark and a signing service as target applications.
\end{itemize}

\section{Background}
\label{sec:background}

\glsreset{CFA}

\glspl{TEE} protect data and processes from high privileged adversaries, an attacker model especially relevant for cloud environments. 
In this section, we discuss
the 
working principles of \glspl{TEE} and one of its examples, Intel 
\gls{SGX}.  
Additionally, we introduce \gls{CFA} and its operations. %

\subsection{Trusted Execution Environments}
\label{sec:background:tees}

In untrusted environments, users face the threat of a high privileged attacker inspecting or modifying their data and processes. 
To protect against such attacks, users can deploy \glspl{TEE}~\cite{suh2003aegis, evtyushkin2014isox, costan2016sanctum, weiser2019timber, lee2020keystone, arm2009trustzone, boivie2013secureblue, costan2016intel,kaplan2016amd, intel2020tdx, hunt2021confidential, arm2021cca}.
\glspl{TEE} aim to ensure the confidentiality and integrity of their data and processes 
even in the presence of a high privileged attacker. 
To defend against such a strong threat, most \glspl{TEE} make use of a combination of hardware and software mechanisms.
These mechanisms include processes which attest the integrity of the \gls{TEE} at launch-time. 
Using these processes, the \gls{TEE}'s owner is able to verify that the \gls{TEE} has been set up correctly before provisioning it with critical data.  

In the last five years, a growing number of cloud providers have incorporated \glspl{TEE} in their infrastructure~\cite{azure2017sgx, ibm2017sgx, alibaba2018sgx, intel2020secustack, google2020confidential}.
At the point of writing, the most prominent example of \glspl{TEE} in cloud environments is Intel \gls{SGX}. 
\gls{SGX} hosts its \glspl{TEE} as userspace applications and splits these applications into an untrusted part, the \emph{\gls{HA}}, and a trusted part, the \emph{enclave}.
To transfer execution into the enclave, the \gls{HA} performs an \emph{ecall}. 
An ecall is similar to a traditional function call, with the main difference being that the function is executed within the protected enclave. 
If the enclave requires assistance from the \gls{HA},
it executes an \emph{ocall}. 
This is, for example necessary to perform I/O operations, as the enclave is not able to directly interact with the operating system.  
When returning from the ocall, the \gls{HA} hands control back to the enclave. 

During launch of an enclave, the \gls{SGX} firmware measures all of the enclave's components. 
This enables a remote user to verify the launch-time integrity of the enclave using \gls{SGX}'s remote attestation process~\cite{anati2013innovative, johnson2016intel, scarlata2018supporting}. 
Once the enclave is launched, firmware and hardware ensure that its memory cannot be accessed by a higher privilege layer, such as the operating system. 

However, the attestation mechanisms of \glspl{TEE} only ensure their integrity at launch-time, not during run-time. 
At run-time, an attacker might be able to extract or modify data from the \gls{TEE} by exploiting vulnerabilities in the code running within the \gls{TEE}. 
An example of such a vulnerability could be a flawed service offered by the \gls{TEE}, enabling an attacker to modify the service's control flow and to execute arbitrary code within the \gls{TEE}. 
Such attacks cannot be detected by the \gls{TEE}'s attestation mechanism, as it only ensures the \gls{TEE}'s integrity during launch. 
It is exactly this gap which we aim to close by combining \glspl{TEE} with \gls{CFA}. 

\subsection{Control-Flow Attestation}
\label{sec:background:cfa}

\begin{figure}[tb]
	\centering
	      \begin{tikzpicture}[font=\footnotesize]
	\tikzset{>=latex}
	
	\node[align=left] at (5.0,7.2) {\normalsize N\textsubscript{1}: \texttt{read input}\\
									\normalsize N\textsubscript{2}: \texttt{compare input to password}\\
									\normalsize N\textsubscript{3}: \quad \texttt{if they match, get private data}\\
									\normalsize N\textsubscript{4}: \quad \texttt{else get public data}\\
									\normalsize N\textsubscript{5}: \texttt{return data}
	};
	
	\node[thick,circle,draw, minimum size=1cm] (1) at (5.0, 6.0) {N\textsubscript{1}};
	\node[thick,circle,draw, minimum size=1cm] (2) at (5.0, 4.6) {N\textsubscript{2}};
	\node[thick,circle,draw, minimum size=1cm] (3) at (3.0, 3.2) {N\textsubscript{3}}; 
	\node[thick,circle,draw, minimum size=1cm] (4) at (7.0, 3.2) {N\textsubscript{4}};
	\node[thick,circle,draw, minimum size=1cm] (5) at (5.0, 1.8) {N\textsubscript{5}};
	\node[thick,circle,draw, minimum size=1cm, dashed] (6) at (7.5, 1.8) {N\textsubscript{6}};
	
	\draw[thick,->] (1) -- (2); 
	\draw[thick,->] (2) -- (4);
	\draw[thick,->] (4) -- (5);
	\draw[->,lightgray] (2) -- (3);
	\draw[->,lightgray] (3) -- (5);
	\draw[->,dashed] (4) .. controls (5.5,3.6) and (4.5,2.4) .. (3);
	\draw[->,dashed] (4) .. controls (8.5, 2.3) .. (6);
	
    \node[devil, mirrored, minimum size=0.75cm] at (7.7,4.1) {};
\end{tikzpicture}
	\caption{An example for legal and illegal control flows in a CFG. 
			 Nodes N\textsubscript{1}--N\textsubscript{5} denote executions of code, while the arrows denote the different edges. 
			 A vulnerability in N\textsubscript{4} can allow an attacker to illegally redirect control flow to N\textsubscript{3} or a new Node N\textsubscript{6}.
	 	 }
	\label{fig:cfg}
\end{figure}
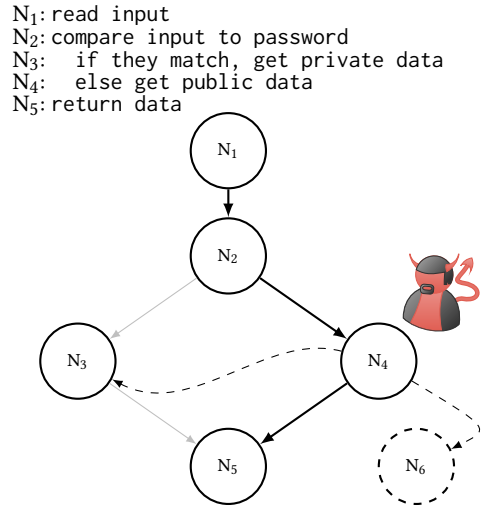

Abera et al.~\cite{abera2016cflat} first introduced 
\gls{CFA} in 2016. 
Using \gls{CFA}, the \emph{prover} provides a second party, the \emph{verifier}, with information about the executed control flow of a program, the \emph{target}.  
Having access to the control flow, the verifier can determine if the target correctly executed the expected sequence of commands. 
To ensure the integrity of the verifier even in case of a compromised \target or prover, verifier and prover are clearly separated.

To perform the verification, the verifier first collects all legal control flows of the \target and calculates its legitimate \gls{CFG}.
In this \gls{CFG}, the target's execution is abstractly represented with \emph{nodes} representing uninterruptible instruction sequences
and \emph{edges} representing transitions between nodes.
\autoref{fig:cfg} shows a \gls{CFG} consisting of the nodes \emph{N\textsubscript{1}} -- \emph{N\textsubscript{5}} and five solid edges marking valid transitions between the nodes. 
In this example, the target first reads input in N\textsubscript{1} and compares it to a stored password in N\textsubscript{2}.
Afterwards, the target will either continue to N\textsubscript{3} if the input and the password matched or to N\textsubscript{4} if the authentication failed. 
Finally, both N\textsubscript{3} and N\textsubscript{4} will hand control to N\textsubscript{5}, which returns the data fetched in N\textsubscript{3} or N\textsubscript{4}. 

To create the \target's \gls{CFG}, the verifier can either make use of static or dynamic analysis.   
With static analysis, the verifier analyzes the binary code of the \target to create the \gls{CFG}. 
However, this approach cannot include edges 
calculated at run-time, such as indirect calls. 
In comparison, dynamic analysis avoids such issues by executing the target to collect all possible paths of the \gls{CFG}. 
For this, the verifier executes the target with different inputs and merges all executed paths to the \gls{CFG}. 
Yet, using this approach, it is difficult to determine the point at which the target executed all possible control flows at least once. 
Therefore, efficiently creating complete \glspl{CFG} for complex programs is still an open research question.
To facilitate the invocation of all control flows, one can resort to programs with reduced complexity.

Now, let us assume that \emph{N\textsubscript{4}} contains a vulnerability which enables an attacker to read and write arbitrary data memory. 
This would allow the attacker to perform a control flow attack and, for example, to modify a return address on the stack.
With the modified return address, the target might then create a new edge by jumping directly from N\textsubscript{4} to N\textsubscript{3}. 
Overwriting the return address would also allow for the creation of a new node N\textsubscript{6}
to perform \gls{ROP}~\cite{shacham2007geometry}.

\gls{CFA} can detect such attacks. %
Once the verifier has calculated the \gls{CFG}, the prover records the execution path of the target and sends it to the verifier. 
By analyzing the execution path, the verifier can determine if the recorded path was within the expected \gls{CFG}. 
If this is not the case, the verifier can conclude that the target has been compromised. 
The verifier can then trace back the \target's execution path to determine the first mismatching edge, allowing to locate the node in which the compromise has first taken effect.

In comparison, \gls{CFI} verifies edges before their execution by comparing them against meta data of permitted backward~\cite{burow19shining} or forward edges~\cite{burow17cfi}. 
To protect this data from attackers, \gls{CFI} commonly relies on information hiding~\cite{zieris18leak}. 
Yet, the past has seen various attacks which were nevertheless able to identify and modify the hidden data, thereby circumventing the \gls{CFI} protection~\cite{gawlik16enabling, goktacs16undermining, oikonomopoulos16poking}.
To defend against such attacks, \gls{CFA} strictly separates prover and verifier, allowing required meta data to be safely stored on the verifier.

\section{Threat Model}
\label{sec:threat}

In this work, we focus on cloud environments, providing us with a virtual machine with full administrative privileges, as it is a common case in such environments. 
This virtual machine runs the service which we aim to protect, our target.
As with most services, we have to assume that the \target's clients are potentially malicious. 
By exploiting vulnerabilities in the \target, a malicious client can gain full control over the \target's memory at a specific point in time. 
While we do assume that the \target is protected by techniques such as Data Execution Prevention, the attacker will still have full control over the \target's data memory. 
Being in control over the data memory, the attacker can perform control flow attacks (\autoref{sec:background:cfa}).

As we are hosting the \target in a cloud environment, we further consider the possibility of a \rogueadmin controlling the software and hardware infrastructure.
Whether this attacker gained access to the infrastructure by escaping a virtual machine~\cite{vmware2017vmsa, ssd2018oracle, citrix2019hypervisor, citrix2020hypervisor} hosted on the same physical system, or the infrastructure has been set up for malicious purposes is of no importance. 
Having control over the infrastructure, the \rogueadmin can extract and modify data and processes within our virtual machine. 
The attacker can either achieve this goal by using software-based methods, such as Virtual Machine Introspection~\cite{garfinkel2003virtual}, or hardware-based methods, such as DMA~\cite{becher2005firewire, boileau2006hit} or cold boot attacks~\cite{halderman2008lest}. 
However, we do not consider 
advanced physical attacks such as tapping buses or dismantling the CPU. 
In other words, we assume the same threat model for a \rogueadmin as \glspl{TEE} such as Intel \gls{SGX}~\cite{costan2016intel}.

\section{Design}
\label{sec:design}

\glsreset{ASLR}

The main goal of \name is to detect control flow attacks in untrusted cloud environments, posing several challenges to overcome. 
In this section, we identify those challenges and present our respective solutions that form the design of \name. 
After giving an overview of our design, we continue with explaining the details of its two phases, the offline and the online phase. 

\subsection{Overview}
\label{sec:design:overview}

In cloud environments, the \target to be attested will likely be a service, such as a signing service for health certificates.   
These services are
deployed within the European Union to sign the so-called Digital Green Certificates providing proof of vaccination against or recovery from COVID-19 or a negative test result~\cite{ehealth2021guidelines}.
To avoid the forging of health certificates, it is important that the service's private key is equally protected from a \rogueadmin and client. %
\name protects the \target signing service against both types of adversaries: 
By using a \gls{TEE}, \name prevents 
an administrator
from accessing the service's private key. 
Additionally, it allows to detect a 
client
compromising the service. 
In the event of such a compromise, the \sprovider can use the collected information to identify and resolve the vulnerability. 
Then, by installing a new private key and revoking health certificates signed with the old key, the impact of the attack
can be minimized. 

Another relevant aspect of signing services is that they only sign newly created health certificates and are therefore only required to provide a limited throughput. 
In other words, their performance is of much lower priority than their security guarantees, which could be 
threatened by a \rogueclient. 
To protect against such a threat, 
we move the attestation away from the client to another, trusted entity. 
To be precise, contrary to previous designs for \gls{CFA}~\cite{abera2016cflat, dessouky2017lofat, zeitouni2017atrium, toffalini2019scarr}, in which the client attests a remote execution, 
we assign this task to the owner of the 
service. 
This separation serves three purposes. 
First, it allows us to leave the interface between the client and the service unchanged, making \name transparent to the client. 
Second, it enables us to perform \gls{CFA} without requiring any information from the client, for example to ensure the freshness of the attestation~\cite{abera2016cflat, dessouky2017lofat, zeitouni2017atrium, toffalini2019scarr, abera2019diat, conti2019radis, sun2020oat}. 
Third, it allows us to perform the attestation time-independent from the communication between client and service.  

Another concern in cloud environments is that we cannot rely on the underlying infrastructure, as it might be controlled by a \rogueadmin.
To protect against this threat, we 
use
\glspl{TEE}.
Specifically, we place both target and verifier in a separate \gls{TEE}, shielding them from attacks by a \rogueadmin.
By using two different \glspl{TEE}, we additionally limit the influence of a \rogueclient on the verifier after exploiting vulnerabilities in the target. 

When choosing a \gls{TEE}, we have to consider that cloud environments are limited in the choice of available hardware. 
Further, we are not able to perform any hardware modifications, requiring us to use the \gls{TEE} offered by the cloud provider. 
Therefore, we discuss \name's design based on the generic concepts of \glspl{TEE}, preserving the flexibility of which \gls{TEE} to choose.  
In \autoref{sec:implementation}, we show how \name can be used in the example of Intel SGX.

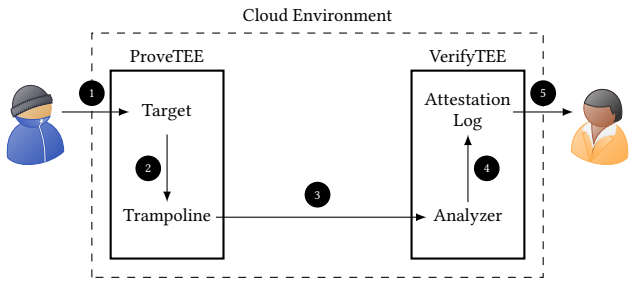
\begin{figure}[t]
	\centering
	\begin{tikzpicture}[font=\footnotesize]
	\tikzset{>=latex}
	
	\draw[dashed] (2,0.5) rectangle (8,3.75);
	\node at (5,4) {Cloud Environment};
	
	\draw[thick] (2.25,0.75) rectangle (3.75,3.25);
	\node [above] at (3,3.25) {ProveTEE};  
    \node at (3,1.3) {Trampoline};  
	\node[text width=1.2cm, align=center] at (3,2.7) {Target};
	
	\draw[thick] (6.25,0.75) rectangle (7.75,3.25);
	\node [above] at (7,3.20) {VerifyTEE};
	\node at (7,1.3) {Analyzer};
	\node[text width=2cm, align=center] at (7,2.7) {Attestation\\Log};
	
	\node[criminal, minimum size=0.75cm] at (1.25,2.5) {};
	\node[alice, mirrored, minimum size=0.75cm] at (8.75,2.5) {};
	
	\draw [->] (1.6,2.7) -- (2.5,2.7);
	\draw [->] (3.0,2.40) -- (3.0,1.5);
	\draw [->] (3.65,1.30) -- (6.45,1.30);
	\draw [->] (7.0,1.5) -- (7.0,2.4);
	\draw [->] (7.6,2.7) -- (8.4,2.7);

	\node[circle,fill=black,text=white,scale=0.7] at (2.0, 2.95) {1};
	\node[circle,fill=black,text=white,scale=0.7] at (2.75, 1.95) {2};
	\node[circle,fill=black,text=white,scale=0.7] at (5.0, 1.60) {3};
	\node[circle,fill=black,text=white,scale=0.7] at (7.25, 1.95) {4};
	\node[circle,fill=black,text=white,scale=0.7] at (8.0, 2.95) {5};
	
\end{tikzpicture}
	\caption{
		The generic design of \name. 
		By using \glspl{TEE}, we prevent a \rogueadmin from extracting or modifying critical data. 
		Additionally, the separation into two \glspl{TEE} ensures that a malicious client is not able to influence \gls{CFA} results by compromising the \target or the \prover. 
	}
	\label{fig:design}
\end{figure}

\autoref{fig:design} depicts the generic design of \name.  
The dashed line indicates the cloud environment offered by the cloud provider. 
Within this environment, we create two \glspl{TEE}, the \emph{\prover} and the \emph{\verifier}. 
The \prover is responsible for hosting the \target. 
Further, it also contains the \emph{\trampoline}, responsible for collecting the control flow executed by the target. 
We insert the \trampoline at compile-time, making it transparent to the client. 
In comparison, the \verifier hosts the \emph{\analyzer}, responsible for verifying the control flow collected by the \trampoline. 
After verification, the \analyzer stores the result in the \emph{\rlog} also located in the \mbox{\verifier}. 
The \verifier itself is entirely passive: 
its tasks are the analysis of control flow data received from the \trampoline and the provision of the \rlog to the \sprovider.
Furthermore, the \verifier may be equipped with additional functionalities such as deploying \target-specific mitigations or shutting down the \target and its service, which are out of the scope of this paper.

Before the \verifier 
is able to 
detect control flow attacks, we require an offline phase in which the \analyzer learns the \target's \gls{CFG} (\autoref{sec:design:offline}).
Afterwards, in the online phase, the client sends regular requests to the \target (\phase{1}). %
While the \target is processing the request, the \trampoline records the control flow (\phase{2}). 
Next, the \trampoline forwards the information to the \analyzer in the \verifier (\phase{3}).  
The forwarding prevents an attacker in control of the \target or even the entire \prover from tampering with the control-flow analysis.   
Having received the information from the \trampoline, the \analyzer compares the collected control flow against the \target's \gls{CFG}. 
Next, it stores the result of this comparison in the \rlog (\phase{4}). 
Finally, the \sprovider is able to retrieve the \rlog from the \verifier (\phase{5}) to perform the deferred attestation of the \target's control flows.

\subsection{Offline Phase}
\label{sec:design:offline}

In the offline phase, we prepare the environment. 
For this preparation, we instrument the \target to regularly pass control to the \trampoline. 
This instrumentation enables the trampoline to record the \target's executed edges 
(\phase{2}) and to provide this information to the \analyzer (\phase{3}).
As we perform the instrumentation at compile-time, we do not require any high-level modifications.
Instead, the developer only needs to annotate where the attestation of the \target's control flow should start and end.   
This makes our instrumentation transparent to the \target's clients, as the interface between the client and the \target remains unchanged. 

Once the \trampoline is able to record the \target's control flow, we calculate the \target's \gls{CFG} by sending requests to the \target until it has executed every control flow path at least once. 
To ensure that no malicious paths are followed, we perform this step in a safe environment without any possibly malicious clients. 
During the dynamic execution, the \analyzer collects all recorded control flows and merges them into the \gls{CFG}.

Yet, determining the point at which a complex \target has executed all legal control flows is still an open research question.  
Therefore, \name aims to protect targets with reduced complexity that are frequently deployed in cloud environments, the so-called microservices. 
Such microservices split the functionality of a complex multi-functional service into multiple, simple services with a single functionality. %
This reduced functionality also reduces the complexity of each service, simplifying the execution of all legal control flows for each of them.  
As an alternative to splitting up the \target itself, we can break down the \gls{CFG} of a complex \target into multiple segments 
to facilitate the collection of each segment's \gls{CFG}, as suggested by previous work~\cite{abera2019diat, toffalini2019scarr}. %

Another difficulty for the creation of the \gls{CFG} is that in cloud environments we likely have to deal with \gls{ASLR}~\cite{spengler2003pax}. 
\gls{ASLR} causes the \target to use different virtual addresses at every launch, preventing us from using addresses to identify the endpoints of edges in the \gls{CFG}.  
Instead, we identify the endpoints by assigning them unique \emph{\callids}, leaving the identification unaffected by ever-changing virtual addresses.

Having determined the \target's full \gls{CFG}, the \analyzer is able to detect control flow attacks on the \target.
Additionally, having access to the full \gls{CFG} enables the \analyzer to identify the node in which the compromise has first shown effect. 

\subsection{Online Phase}
\label{sec:design:online}

During the online phase, the \target hosted in the \prover waits for requests from potentially malicious clients (\phase{1}).
As \name is transparent to the clients, they can send the same requests to the instrumented \target as to the original. 

While the \target processes the request, it records all \callids 
and provides them to the \trampoline (\phase{2}).
Yet,
the \trampoline itself is not responsible for processing the IDs. 
This is due to the fact that it runs in the \prover, next to the \target. 
Hence,
an attacker gaining control over the \target 
could also
tamper with the \trampoline. 
To ensure that such an attack does not influence the \gls{CFA},
we outsource the analysis task to the \verifier. 
Specifically, the only task of the \trampoline is to forward the collected IDs to the \analyzer in the \verifier (\phase{3}).   
This approach has two advantages. 
First, it reduces the \trampoline's processing time. 
And second, it protects the analysis 
from an attacker in control of the \target and the \prover. 

Next, the \analyzer uses the received \callids to reconstruct the 
\target's control flow.
By then comparing the 
control flow against the \target's \gls{CFG} created in the offline phase (\autoref{sec:design:offline}), the \analyzer 
can
detect control flow attacks.
After
the control-flow attestation,  
the \analyzer stores 
the respective result 
in the \rlog (\phase{4}).

Finally, the \sprovider fetches the \rlog from the \verifier (\phase{5}).
This log contains all requests which altered the \target's control flow, allowing for a deferred attestation of the \target. 
When such requests have been logged, the \sprovider can take appropriate actions. 
In the example of our signing service, this would include the revocation of erroneously signed certificates as well as analyzing and resolving bugs in the \target's code.

\section{Implementation}
\label{sec:implementation}

\glsreset{HA}

Based on our design, we implemented the \name prototype using Intel SGX (\autoref{sec:background:tees}) as \gls{TEE}.
The reason for this choice is that SGX is supported by most modern Intel CPUs and is also widely available in cloud environments~\cite{azure2017sgx, ibm2017sgx, alibaba2018sgx, intel2020secustack}.
In this section, we discuss implementation details of the prototype, such as the instrumentation of the target and the distribution of work between different threads.   

\subsection{Overview}
\label{sec:implementation:overview}

\begin{figure}[t]
	\centering
	\begin{tikzpicture}[font=\footnotesize]
	\tikzset{>=latex}
	
	\draw[dashed] (2,0.5) rectangle (8,3.75);
	\node at (5,4) {Cloud Environment};
	
	\draw[thick] (2.25,0.75) rectangle (3.75,3.25);
	\node [above] at (3,3.25) {ProveTEE};  
	\node at (3,1.3) {Trampoline};  
	\node[text width=1.2cm, align=center] at (3,2.7) {Target};
	
	\draw[thick] (6.25,0.75) rectangle (7.75,3.25);
	\node [above] at (7,3.20) {VerifyTEE};
	\node at (7,1.3) {Analyzer};
	\node[text width=2cm, align=center] at (7,2.7) {Attestation\\Log};
	
	\node[criminal, minimum size=0.75cm] at (1.25,2.5) {};
	\node[alice, mirrored, minimum size=0.75cm] at (8.75,2.5) {};
	
	\draw [->] (1.6,2.7) -- (2.5,2.7);
	\draw [->] (3.0,2.40) -- (3.0,1.5);
	\draw [->] (3.70,1.30) -- (4.45,1.30);
	\draw [->] (5.60,1.30) -- (6.45,1.30);
	\draw [->] (7.0,1.5) -- (7.0,2.4);
	\draw [->] (7.6,2.7) -- (8.4,2.7);

	\node[circle,fill=black,text=white,scale=0.7] at (2.0, 2.95) {1};
	\node[circle,fill=black,text=white,scale=0.7] at (2.75, 1.95) {2};
	\node[circle,fill=black,text=white,scale=0.7] at (5.0, 1.65) {3};
	\node[circle,fill=black,text=white,scale=0.7] at (7.25, 1.95) {4};
	\node[circle,fill=black,text=white,scale=0.7] at (8.0, 2.95) {5};
	
    \draw[thick] (4.25, 0.75) rectangle (5.75,2.05);
    \node[above, align=center] at (5,2.00) {Shared\ \\Memory};  
    \draw[] (4.50,1.20) rectangle (4.70,1.40);
    \draw[] (4.70,1.20) rectangle (4.90,1.40);
    \draw[] (4.90,1.20) rectangle (5.10,1.40);
    \draw[] (5.10,1.20) rectangle (5.30,1.40);
    \draw[] (5.30,1.20) rectangle (5.50,1.40);
	
\end{tikzpicture}
	\caption{
		Overview of our prototype based on the \name design. 
		The additional shared memory region allows to transfer the IDs collected by the \trampoline to the \analyzer in the \verifier.
	}
	\label{fig:implementation}
\end{figure}
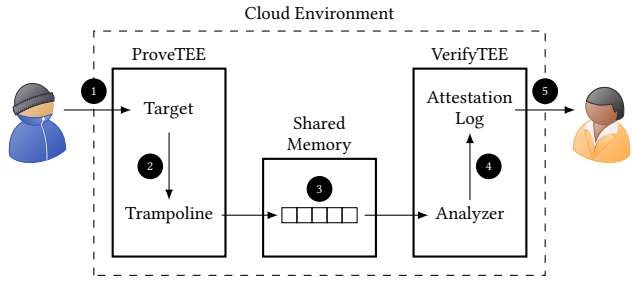

\autoref{fig:implementation} gives an overview of our implementation. 
The basic building blocks are the same as in the design (\autoref{sec:design}), except the additional shared memory region used to exchange data between the \prover and the \verifier.
This region allows the \glspl{TEE} to exchange information without requiring any context switches, thereby reducing the performance overhead of the communication. 

To create the shared memory region, we make use of the fact that SGX \glspl{TEE} share the virtual address space with their \gls{HA}~\cite{schwarz2019practical}. 
Specifically, by using the same \gls{HA} to launch both the \prover and the \verifier, both \glspl{TEE} will share the same virtual address space~\cite{morbitzer2019scanclave}.
This enables us to allocate a memory region in the \gls{HA} which both the \prover and the \verifier can access, allowing them to exchange collected IDs.
Note that while both \glspl{TEE} are able to read and write the shared memory, they are not able to access each other's memory.  

The first step of our \name prototype is to perform an offline phase, in which it collects all of the \target's control flows to calculate its \gls{CFG} (\autoref{sec:design:offline}). 
Afterwards, our prototype verifies the \target's control flow in the online phase and provides the verification results to the \sprovider (\autoref{sec:design:online}). 

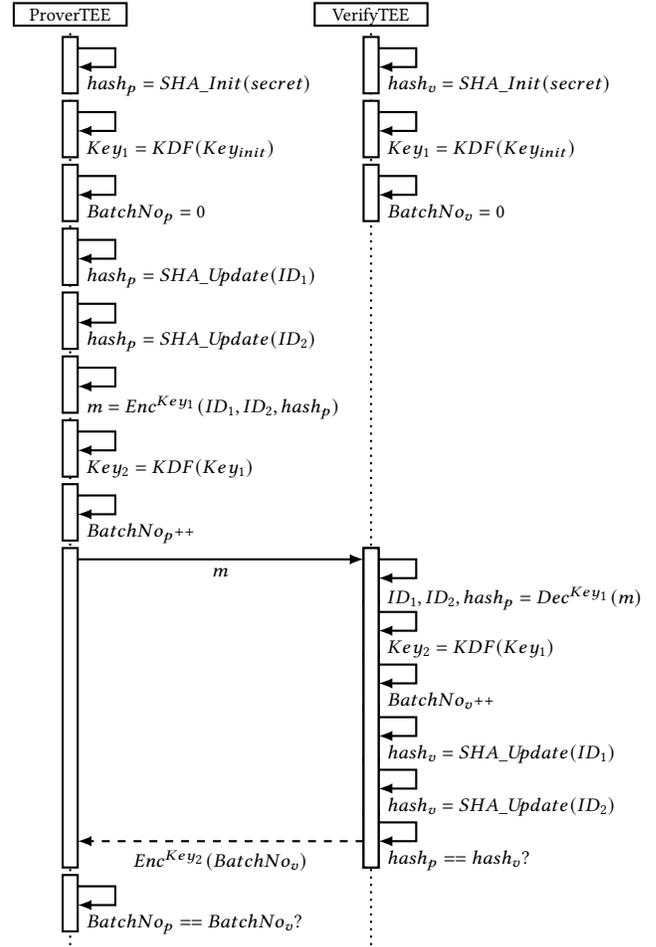
\begin{figure}[t]
	\centering
	\newcommand{\action}[3]{
	\draw[thick, fill=white] (#1,#2) rectangle (#1+0.2,#2+0.75);
	\draw[thick,->] (#1+0.2,#2+0.6) -- (#1+0.7,#2+0.6) -- (#1+0.7,#2+0.35) -- (#1+0.2,#2+0.35) node[below right] {#3};
}

\begin{tikzpicture}[font=\footnotesize]
	\tikzset{>=latex}

	\draw[thick] (0.25,11.0) rectangle (1.75,11.3);
	\node [] at (1,11.15) {ProverTEE};
	\draw[thick, dotted] (1,10.9) -- (1,-1.3);

	\draw[thick] (4.25,11.0) rectangle (5.75,11.3);
	\node[] at (5,11.125) {VerifyTEE};
	\draw[thick, dotted] (5,10.9) -- (5,-1.3);
	
	\action{0.9}{10.1}{$hash_p = SHA\_Init(secret)$};
	\action{4.9}{10.1}{$hash_v = SHA\_Init(secret)$};

	\action{0.9}{9.25}{$Key_1 = KDF(Key_{init})$};
	\action{4.9}{9.25}{$Key_1 = KDF(Key_{init})$};
	
	\action{0.9}{8.40}{$BatchNo_p = 0$};
	\action{4.9}{8.40}{$BatchNo_v = 0$};	
	
	\action{0.9}{7.55}{$hash_p = SHA\_U$\hspace{-0.06cm}$pdate(ID_1)$};
	\action{0.9}{6.7}{$hash_p = SHA\_U$\hspace{-0.06cm}$pdate(ID_2)$};

	\action{0.9}{5.85}{$m = Enc^{Key_1}(ID_1,ID_2,hash_p)$};	
	\action{0.9}{5.0}{$Key_2 = KDF(Key_1)$};
	\action{0.9}{4.15}{$BatchNo_p$++};
	
	\action{4.9}{3.3}{$ID_1,ID_2,hash_p = Dec^{Key_1}(m)$};
	\action{4.9}{2.60}{$Key_2 = KDF(Key_1)$};	
	\action{4.9}{1.90}{$BatchNo_v$++}
	\action{4.9}{1.20}{$hash_v = SHA\_U$\hspace{-0.06cm}$pdate(ID_1)$};
	\action{4.9}{0.5}{$hash_v = SHA\_U$\hspace{-0.06cm}$pdate(ID_2)$};
	\action{4.9}{-0.2}{$hash_p == hash_v?$};	
	\draw[thick, fill=white] (0.9,-0.2) rectangle (1.1,4.05);	
	\draw[thick,->] (1.1,3.9) -- (4.9,3.9) node[midway,below] {$m$};
	\draw[thick, fill=white] (4.9,-0.2) rectangle (5.1,4.05);	
	\draw[thick,->, dashed] (4.9,0.15) -- (1.1,0.15) node[midway,below] {$Enc^{Key_2}(BatchNo_v)$};

	\action{0.9}{-1.05}{$BatchNo_p == BatchNo_v?$};
	
\end{tikzpicture}
	\caption{
		The exchange of IDs between the \prover and the \verifier. 
		To ensure the integrity of IDs cached in the \batches, we make use of a hash chain. 
		Additionally, we calculate a new encryption key for every exchange with the help of a KDF and record the number of the current batch. 
	}
	\label{fig:seq}
\end{figure}

In the implementation, each basic block in the \target corresponds to a node in the \gls{CFG}. 
The basic blocks are connected via edges, the endpoints of which are recorded by the \trampoline. %
To be precise, the \target calls the \trampoline each time before entering or exiting a basic block, providing the \callid of the respective endpoint.  
The \trampoline stores this ID in an \emph{\batch}, which acts as a cache. 

\autoref{fig:seq} depicts how we are able to securely cache the IDs by combining them with the calculation of a new key for every data exchange and tracking the number of exchanged \batches. 
To ensure the integrity of the cached IDs in the \prover, we make use of a hash chain.
The root of this hash chain is a secret we provide to both the \prover and the \verifier. 
When we cache an ID in the \batch, we update the hash in the \prover, $hash_p$, with the new ID. 
If the \batch is full, we combine it with the current value of $hash_p$ to form a single message, which we then encrypt and store in the \emph{\queue} located in the shared memory region.
After the \verifier reads and decrypts the message from the \queue, it adds all IDs to its own hash, $hash_v$. 
As we initialize both $hash_p$ and $hash_v$ with the same secret and add the same IDs to both hashes, $hash_v$ should equal the value of $hash_p$ transmitted with the batch.
To minimize the performance impact of this step, our prototype uses BLAKE3 as hash algorithm~\cite{oconnor2021blake3}.

When exchanging the \batch via shared memory, we have to keep in mind that the memory is also accessible to other, untrusted entities such as the \gls{HA} or the operating system.  
This means that a \rogueadmin would be able to inspect and modify transferred \batches.
To protect the batches from such attacks, we implement mechanisms which ensure their confidentiality and integrity. 
For confidentiality, we provide both \glspl{TEE} with an initial encryption key, $Key_{Init}$. 
Both \glspl{TEE} use this initial key as input to a \gls{KDF} 
to calculate an encryption key.
For the following messages, the encryption key serves as input for the \gls{KDF} to produce a new key for the next message, allowing us to calculate a new key for each message. 
These keys enable us to encrypt the \batches and the respective hash values before writing them into shared memory, ensuring that they cannot be inspected by a \rogueadmin.  
By additionally calculating a new key for every message, we ensure that an attacker in control of the \prover 
cannot
infer previously used keys. 
We achieve this by calculating the keys with an irreversible \gls{KDF} and by deleting old keys after usage.  
To be precise, we again make use of BLAKE3, as it is irreversible and already integrated in our prototype. 

To ensure the integrity of the communication, we use AES-GCM for encryption, which additionally provides integrity protection~\cite{mcgrew2004galois}.
AES-GCM requires an IV, which we need to synchronize between the \trampoline and the \analyzer to ensure correct encryption and decryption. 
We achieve this synchronization by managing an independent counter, \emph{BatchNo}, in both the \prover and the \verifier. 
After each encryption or decryption, we increase the counter to keep both counters synchronized. 
These counters allow us to detect attacks in which the \rogueadmin prevents forwarding of the \batches. 
For this detection, the \verifier regularly acknowledges the receipt of the \batches. 
On the other side, the \prover, having stored a certain number of \batches into the \queue,  will wait for an acknowledgment before continuing to execute the \target. 
We define the frequency of these acknowledgments as the \emph{\ff}.
In \autoref{sec:evaluation}, we analyze the performance and security impact of different feedback frequencies. 

\subsection{Instrumentation}
\label{sec:implementation:instrumentation}

For the automatic instrumentation of the \target, we rely on the LLVM compiler framework.
The instrumentation allows us to establish the \target's \gls{CFG} in the offline phase (\autoref{sec:design:offline}) and to verify its control flows in the online phase (\autoref{sec:design:online}).

To calculate the \gls{CFG}, we need to determine all edges executed by the \target. 
Each edge has two endpoints: the exit from a basic block and the entry into another basic block, both of which we identify with a unique \callid.
By analyzing the recorded \callids, we 
can
reconstruct all edges executed by the \target and therefore its full control flow.  
To collect the \callids, we add calls to the trampoline on every entry and exit of each basic block. 

To instrument all entries and exits, we modify two different phases of the LLVM compilation process: the IR optimization and the backend. 
In our IR pass, we add calls to the \trampoline at the beginning and the end of each basic block and before and after direct function calls.
To assign a unique \callid for each endpoint, we 
make use of 
a counter. 

In our backend pass, we instrument all indirect branches. 
Specifically, we add calls to the trampoline before and after every indirect function call, before every indirect jump, and before every return instruction. 
As before, each trampoline call is associated with a unique \callid, allowing us to identify the endpoints of all executed edges in the \target's control flow. 
Additionally, on indirect calls, indirect jumps, and returns we XOR the \callid with the offset between the current instruction pointer and the jump destination to detect modifications of the jump address.
This safeguard allows us to detect jumps from instrumented into uninstrumented code. 
Specifically, it causes a modification of the jump destination to also change the recorded \callid, allowing us to detect the modification even if it points the execution to uninstrumented code. 
Being able to record all control flows, we use the offline phase to combine the collected \callids to the \target's \gls{CFG}.

While it may technically be possible to perform all instrumentations in IR only, the backend pass allows us to precisely place the calls to the \trampoline. 
For example, let us consider \gls{ROP} attacks in which 
gadgets consist
of the last instructions before a return instruction~\cite{shacham2007geometry}. 
To detect such attacks, we need to place the call to the trampoline as close to the return instruction as possible. 
When adding the call in IR, the subsequent code generation in the backend will insert instructions for stack cleanup between the call and the return instruction. 
This would allow 
execution of these instructions
without invoking the trampoline, providing a potential ROP gadget. %
In comparison, using our backend pass, inserting the call during code generation allows 
us to place the call as close to the return instruction, or any other indirect branch instruction, as possible.
Note that we do not consider the instructions before direct function calls as possible \gls{ROP} gadgets due to the hard-coded function address. 
Hence,
we refrain from instrumenting direct function calls in the backend pass and instead instrument them in our IR pass.

\subsection{Execution}
\label{sec:implementation:execution}

The instrumentation of the \target allows us to use dynamic execution to record the \callids 
in the offline phase (\autoref{sec:design:offline}). 
For this, we created a script which performs a variety of requests, causing the \target to execute each valid control flow at least once. 
During processing of these requests, the \analyzer investigates the \callids and creates the target's \gls{CFG}.
We then switch to the online phase, in which the \analyzer uses the \gls{CFG} to detect control flow attacks. 

In the online phase, the \name prototype performs \gls{CFA} for all control flows between a start and end point annotated by the developer. 
Using for example our signing service, we annotated the initialization and the termination of an incoming connection. %
The annotations cause the \trampoline to send a special begin or end tag to the \analyzer, triggering it to start or stop the analysis.
This additional analysis task unavoidably slows down the processing time of a request.
To minimize this impact, we make use of the high processing power of cloud systems. 
Specifically, we provision our virtual machine with multiple vCPUs, thus supporting parallel processing of different tasks. 
To
fully capitalize on this ability, we distribute the prototype's tasks between two threads: the \emph{\pthread} and the \emph{\vthread}. 
   
The \pthread is responsible for executing the target to process incoming requests. 
Additionally, it executes the \trampoline to collect the IDs in batches and to store them in the shared \queue.     
In detail, within the \pthread, the \target waits for and processes incoming requests (\phase{1}). 
During processing, the thread jumps to the \trampoline on every instrumentation (\phase{2}), which caches the recorded ID in the \batch and extends the hash chain. 
When an \batch is full, the \trampoline encrypts it with AES-GCM, using the key derived with the \gls{KDF} (\autoref{sec:implementation:overview}). 
Afterwards, it stores the encrypted \batch in the \queue (\phase{3}). 
When the \trampoline finished its execution, the \pthread returns to the \target to continue processing the request.

In comparison, the \vthread is responsible for the reconstruction of the \target's control flow. 
It reads the IDs provided by the \pthread, validates them against the \gls{CFG} learned in the offline phase, and stores the validation result in the \rlog. %
To fulfill these tasks, the \vthread reads the encrypted \batches from the \queue and decrypts them with the key derived from the \gls{KDF} (\autoref{sec:design:overview}). 
Next, it forwards the IDs contained in the \batches to the \analyzer.
Using the \callids, the \analyzer rebuilds the target's control flow and verifies it against the \target's \gls{CFG}. 
After the \analyzer finished this verification, the \vthread stores the respective result in the \rlog (\phase{4}). 
Afterwards, it checks if new \batches exist in the \queue to continue the attestation. 

The most efficient method to synchronize our two threads would be a mutex~\cite{buttlar1996pthreads}. 
Unfortunately, \gls{SGX} does not yet support the use of mutexes to synchronize multiple threads running in different enclaves. 
Instead, our prototype currently has to rely on atomic operations~\cite{gcc2021atomic} to add or read \batches to or from the \queue. 
Additionally, we use a spinlock~\cite{wehrle2004linux} while the \vthread waits for the \pthread to add new \batches to an empty \queue.

To retrieve the verification results (\phase{5}), \name allows the \sprovider to access the \rlog in a protected manner.
For this access, the \sprovider confirms the integrity of the \verifier using remote attestation~\cite{anati2013innovative, johnson2016intel, scarlata2018supporting}.
This attestation process establishes a TLS connection, thereby creating an encrypted channel directly into the \gls{TEE}~\cite{knauth2018integrating}. 
Using this encrypted channel, the \sprovider is able to securely retrieve the \rlog. 

To summarize, the \analyzer uses the offline phase to calculate the \target's \gls{CFG}. %
In the online phase, it verifies the \target's recorded control flow by comparing the \callids with the previously constructed \gls{CFG}. 
To reduce the impact of this analysis on the \target's processing time, we split the tasks between two threads, the \pthread and the \vthread. 
While the \pthread executes the \target and collects its control flow information, the \vthread analyses the collected information.

\section{Evaluation}
\label{sec:evaluation}

To 
precisely evaluate the overhead of the different steps of our design, we analyzed the processing times of the different components (\phase{2} -- \phase{4}) individually.
Additionally, we evaluated the performance overhead and the security of the entire prototype. 

\subsection{Setup}
\label{sec:evaluation:setup}

As \target for the evaluation, we created a microservice responsible for signing health certificates (\autoref{sec:design:overview}). 
Our signing service receives the hash of a newly created health certificate, signs the hash with its private key, and returns the signature. 
The service performs this exchange via a secure connection, which we establish based on code from the SGX-OpenSSL project~\cite{juhyeng2020sgxopenssl}.
This code allows clients to establish a TLS channel directly into the \gls{TEE}. 
By using annotations, we instructed \name to perform the \gls{CFA} between receiving and terminating an incoming connection.  
In total, our signature service consists of $4{,}533$ instructions, excluding libraries such as OpenSSL, and the \trampoline of $47{,}542$ instructions. 

As traditional time measurements are not available within SGX \glspl{TEE}~\cite{brasser2019drsgx}, we added an ocall to both \glspl{TEE} which notifies the \gls{HA} to start or stop a measurement. 
By executing this ocall at the start and at the end of our measurements, we were able to determine the difference between those two points in time in the \gls{HA}. 

We performed our measurements in the Microsoft Azure cloud environment on a Standard DC4s\_v2 machine with an Intel Xeon E-2288G CPU, four vCPUs, and $16$~GiB of memory. 
Within the virtual machine, we were running the default Ubuntu $18.04$ provided by Azure and the Linux \gls{SGX} DCAP driver in version $v1.41$.

\subsection{Benchmark Performance Evaluation}
\label{sec:evaluation:bperformance}

To determine the throughput of \name in extreme conditions, we deployed a benchmark as the \target before evaluating the performance of our signing service. %
We used the \texttt{sgx-nbench} benchmark suite~\cite{utds3lab2021sgxnbench}, which is a port of \texttt{nbench}~\cite{petabridge2020nbench} to \gls{SGX}.
As a baseline, we ran the benchmark without instrumentation.
Then, we instrumented the \target with \name and used our default configuration of an \bsize of $10{,}000$ and a \ff of $10$. 
In Sections \ref{sec:evaluation:cperformance} and \ref{sec:evaluation:pperformance}, we analyze the impact of different batch sizes and feedback frequencies. 

\begin{table}[h]
	\centering
	\caption{
		Comparison of \texttt{sgx-nbench} with and without \name. Measurements are in iterations per second. 
	}
	\label{tab:evaluation:benchmark}
	\begin{tabular}{lrrr}
		\textbf{Test}	& \textbf{Baseline} & \textbf{\name}	& \textbf{Overhead}\\
		\midrule
		Numeric Sort	& $2{,}387.2$		& $11.4$			& $208$x\\
		String Sort		& $1{,}285.1$		& $42.2$			& $29$x\\
		Bitfield		& $9.3382*10^8$		& $3.0783*10^6$		& $302$x\\
		FP Emulation	& $1{,}251$			& $12$				& $103$x\\
		Fourier			& $62{,}574$		& $11{,}438$		& $4$x\\
		Assignment		& $121.51$			& $0.17$			& $713$x\\
		Idea			& $21{,}500$		& $55$				& $390$x\\
		Huffman			& $6{,}509.4$		& $49.3$			& $131$x\\														
		Neural Net		& $192.77$			& $1.00$			& $192$x\\
		LU Decomp.		& $4302.4$			& $23.5$			& $182$x\\
		Average			& -					& -					& $225$x\\
	\end{tabular}
\end{table}

\autoref{tab:evaluation:benchmark} shows the overhead of \name for the \texttt{sgx-nbench} benchmark suite.
The overhead varies between $4$x for the Fourier Test and $713$x for the Assignment Test. 
On average, we recorded an overhead of $225$x between the baseline and the instrumented benchmarks. 
Having determined this relatively high overhead, it is important to note that in comparison to most targets, a benchmark evaluates the performance under extreme conditions, such as a very high CPU load. 
In contrast, a regular \target, such as our signing service, will also perform less resource-intensive operations such as waiting for the operating system or for incoming network traffic. 
This will reduce the overhead of a regular \target in comparison to the evaluated benchmark. 
To prove this claim, we continue by evaluating \name's different components using our signing service as the \target.

\subsection{Component Performance Evaluation}  
\label{sec:evaluation:cperformance}

Using our signing service (\autoref{sec:evaluation:setup}), we continued by evaluating the different components of \name, namely the instrumented \target (\phase{2}), the \queue (\phase{3}), and the \analyzer (\phase{4}). 
To evaluate the throughput of the instrumented target (\phase{2}), we simulated $100{,}000$ client requests. 
While processing the requests, the \target produced $9.87$ IDs per microsecond, or, in other words, called the \trampoline on average every $101.29$~ns. 
We call this frequency, with which the \target calls the trampoline, the \emph{trampoline call frequency}.

Next, we determined the throughput of the \queue used to transfer IDs from the \trampoline to the \analyzer (\phase{3}). 
For this transfer, the \trampoline caches the received IDs in batches (\autoref{sec:implementation:overview}). 
When an \batch is full, the \trampoline encrypts and stores it into the queue, where the \analyzer is able to read it. 
Having received and successfully decrypted the \batch, the \analyzer gives feedback to the \trampoline.
Determined by the \ff (\autoref{sec:implementation:overview}), the \analyzer can be configured to give feedback only after a certain amount of \batches has been processed.
Using this approach, the throughput with which we can transfer the IDs from the \trampoline to the \analyzer depends on two variables: the \batch size and the \ff.  

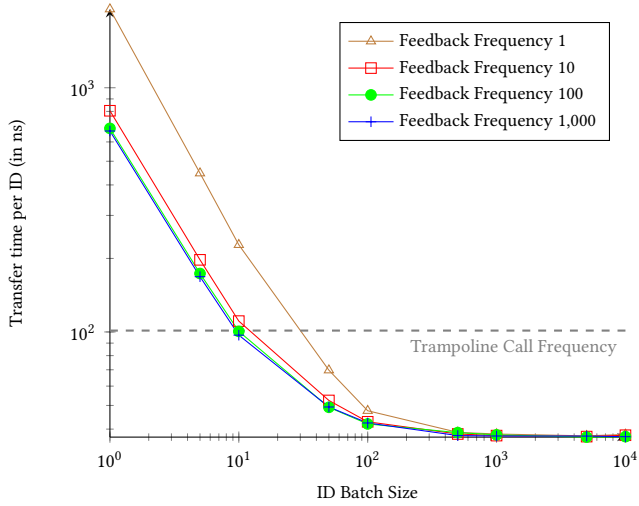
\begin{figure}[t]
	\centering
	\begin{tikzpicture}[font=\footnotesize]
	\tikzset{>=latex}
	\begin{axis}[
		xmode=log, log basis x=10,
		ymode=log,log basis y=10,
		xmin=0,     %
		ylabel=Transfer time per ID (in ns),
		xlabel=ID Batch Size,
		tick align=outside,
		axis lines = left,
		legend pos=north east,
		legend cell align={left}, %
		legend entries={
			Feedback Frequency 1,
			Feedback Frequency 10,
			Feedback Frequency 100,
			Feedback Frequency 1{,}000,
}
		]
		
		\addplot [color=brown, mark=triangle] 
		coordinates {
			(1,2098.106576)(5,446.720155)(10,227.398085)(50,69.721137)(100,47.569346)(500,38.801903)(1000,38.198997)(5000,37.543732)(10000,38.047524)
		};
	
		\addplot [color=red, mark=square] 
		coordinates {
			(1,804.893656)(5,197.409286)(10,110.771623)(50,52.409851)(100,42.860599)(500,38.214251)(1000,37.634240)(5000,37.275859)(10000,37.767088)
		};
	
		\addplot [color=green, mark=*] 
		coordinates {
			(1,681.746762)(5,173.518701)(10,100.524896)(50,49.127742)(100,42.004409)(500,38.706817)(1000,37.896915)(5000,37.102037)(10000,37.264040)
		};

		\addplot [color=blue, mark=+] 
		coordinates {
			(1,665.170681)(5,168.401202)(10,97.016750)(50,49.284134)(100,42.387736)(500,37.646828)(1000,37.487768)(5000,37.462733)(10000,37.252318)
		};
	
		\coordinate (a) at (axis cs:1,101.29);
		\coordinate (b) at (axis cs:10000,101.29);
		\draw[gray, dashed, thick] (a) -- (b); 
		\node[gray, below left] at (b) {Trampoline Call Frequency};
		
	\end{axis}
\end{tikzpicture}
	\caption{
		Average time required to transfer a single ID from the \trampoline to the \analyzer (\phase{3}). 
		While the impact of the \ff is limited, the transfer time drastically decreases with higher batch sizes.
	}
	\label{fig:eval:queue}
\end{figure}

\autoref{fig:eval:queue} gives an overview of the throughput with different \batch sizes and feedback frequencies.
While the X-axis depicts the different batch sizes, the Y-axis indicates the average transfer time per ID in nanoseconds.
Both axes are in logarithmic scale. 
The four different plots indicate the impact of different feedback frequencies on the transfer time. 
A \ff of $1$ means that the \trampoline waits for feedback from the \analyzer after every single \batch. 
In comparison, using a frequency of $1{,}000$, the \trampoline only waits for feedback after having transferred $1{,}000$ \batches. 

The four different plots show that the feedback frequency has only limited impact on the throughput. 
While a frequency of $1$ does decrease the throughput, the differences between the frequencies $10$, $100$, and $1{,}000$ are only minor. 
This difference becomes negligible in combination with an \bsize of $100$ or more.
In comparison, the \bsize has a significantly higher impact on the throughput. 
Using a size of $1$, we needed on average $2{,}098.11$~ns to transfer a single ID with \ff $1$, and $665.17$~ns with a frequency of $1{,}000$. 
Yet, we are able to drastically reduce the transfer time by increasing the \bsize. %
For example, our default configuration of an \bsize of $10{,}000$ and a \ff of $10$ reduces the transfer time to $37.76$~ns.

For a
better interpretation of these results, the dashed gray line in \autoref{fig:eval:queue} indicates the trampoline call frequency, which we previously determined to be $101.29$~ns. 
All configurations achieving transfer times below this frequency transfer the IDs to the analyzer faster than the \target creates new IDs.  
With our prototype, we stay above this frequency for \batch sizes below $10$. 
Yet, a size of $10$ 
combined
with a feedback frequency of $100$ or $1{,}000$ already transfers the IDs faster than the trampoline call frequency. 
Additionally, the transfer time stayed below the trampoline call frequency with an \bsize of $50$ and a feedback frequency of $10$ and $1$.

Having evaluated the throughput of the \target and the \queue, we continued by evaluating the \analyzer's throughput (\phase{4}). 
As we consider the offline phase less time-critical, we focused on the evaluation of the online phase. 
In the online phase, the \analyzer verifies the \target's control flow by comparing the \callids with the previously recorded \gls{CFG}. 
Specifically, it verifies all IDs received between the start and end tag (\autoref{sec:implementation:execution}). 
With our signing service, the \target sends the start tag 
when
the worker thread receives an incoming connection, and the end tag when terminating the connection.
This allows us to only attest handling of a request and to exclude other tasks such as waiting for a new connection. 

To
determine the 
throughput of the \analyzer,
we measured the time required to process the IDs of $100{,}000$ requests. 
For each request, the \analyzer processed around $539$ IDs, for which it required on average $7.34$~ns per ID.
This is significantly lower than the \target's trampoline call frequency of $101.29$~ns. 
In other words, the \analyzer 
processes 
incoming IDs faster than they are produced by the \target.

\subsection{Signing Service Performance Evaluation}  
\label{sec:evaluation:pperformance}

Having evaluated the throughput of \name's components, we continued with evaluating its overall overhead on our signing service. 
To quantify this overhead, we prepared a \gls{TEE} running the uninstrumented service and measured the time required 
to process an incoming request.
Using this setup, we performed $100{,}000$ valid signing requests, which required on average $54.79$~$\mu$s 
 per request.

\begin{figure*}[t]
	\centering
	\begin{tikzpicture}[font=\footnotesize]
	\tikzset{>=latex}
	\begin{axis}[
		xmode=log, log basis x=10,
		ymode=log,log basis y=10,
		xmin=0,     %
		ymin=0,     %
		ymax=3595,   %
		ylabel=Processing Time per Request (in $\mu$s),
		xlabel=ID Batch Size,
		tick align=outside,
		axis lines = left,
		legend pos=north east,
	    legend cell align={left}, %
   		width=.5\textwidth
	]

		\addplot [color=brown, mark=triangle] 
		coordinates {
			(1,755.839863)(5,202.936413)(10,132.642371)(50,73.415195)(100,67.450687)(500,62.110712)(1000,60.821071)(5000,58.557989)(10000,58.096868)
		};

		\addplot [color=red, mark=square] 
		coordinates {
			(1,716.033471)(5,188.625030)(10,120.707308)(50,70.225819)(100,62.070857)(500,58.624768)(1000,58.141205)(5000,57.886363)(10000,57.279583)
		};

		\addplot [color=green, mark=*] 
		coordinates {
			(1,676.005963)(5,135.118679)(10,82.520896)(50,60.222267)(100,58.828707)(500,58.265040)(1000,57.992973)(5000,57.768286)(10000,57.474274)
		};

		\addplot [color=blue, mark=+] 
		coordinates {
			(1,304.163608)(5,78.294424)(10,66.466221)(50,60.062543)(100,58.237934)(500,58.128790)(1000,58.122627)(5000,58.433549)(10000,57.678113)
		};
		
		\coordinate (a) at (axis cs:1,10.66);
		\coordinate (b) at (axis cs:100000,10.66);
		\draw[gray, dashed, thick] (a) -- (b) node[midway, below] {Original processing time};

	\end{axis}
\end{tikzpicture}
\begin{tikzpicture}[font=\footnotesize]
	\tikzset{>=latex}
	\begin{axis}[
		xmode=log, log basis x=10,
		ymode=log,log basis y=10,
		xmin=0,     %
		ymin=0,     %
		ymax=3595,   %
		xlabel=ID Batch Size,
		tick align=outside,
		axis lines = left,
		legend pos=north east,
		width=.5\textwidth,
		legend cell align={left}, %
		legend entries={
			Feedback Frequency 1,
			Feedback Frequency 10,
			Feedback Frequency 100,
			Feedback Frequency 1{,}000,
		}
		]
		
		\addplot [color=brown, mark=triangle] 
		coordinates {
			(1,3594.445080)(5,785.722978)(10,410.742549)(50,113.789622)(100,78.728546)(500,48.382089)(1000,42.894914)(5000,30.625043)(10000,28.809712)
		};
		
		\addplot [color=red, mark=square] 
		coordinates {
			(1,3375.167938)(5,753.524640)(10,392.537151)(50,113.421813)(100,76.919432)(500,47.631431)(1000,41.819980)(5000,30.438115)(10000,28.462842)
		};
		
		\addplot [color=green, mark=*] 
		coordinates {
			(1,3401.596717)(5,743.837422)(10,393.124032)(50,112.867094)(100,76.480727)(500,47.626043)(1000,42.058872)(5000,30.652178)(10000,28.619903)
		};
		
		\addplot [color=blue, mark=+] 
		coordinates {
			(1,3554.689015)(5,749.642823)(10,390.654508)(50,114.847395)(100,77.836749)(500,47.554928)(1000,42.011656)(5000,30.673509)(10000,28.616564)
		};

	\end{axis}
\end{tikzpicture}
	\caption{
		Average processing time for a single request per thread.   
		The left graph depicts the measurements for the \pthread and the right graph for the \vthread.
	}
	\label{fig:eval:processing}
\end{figure*}
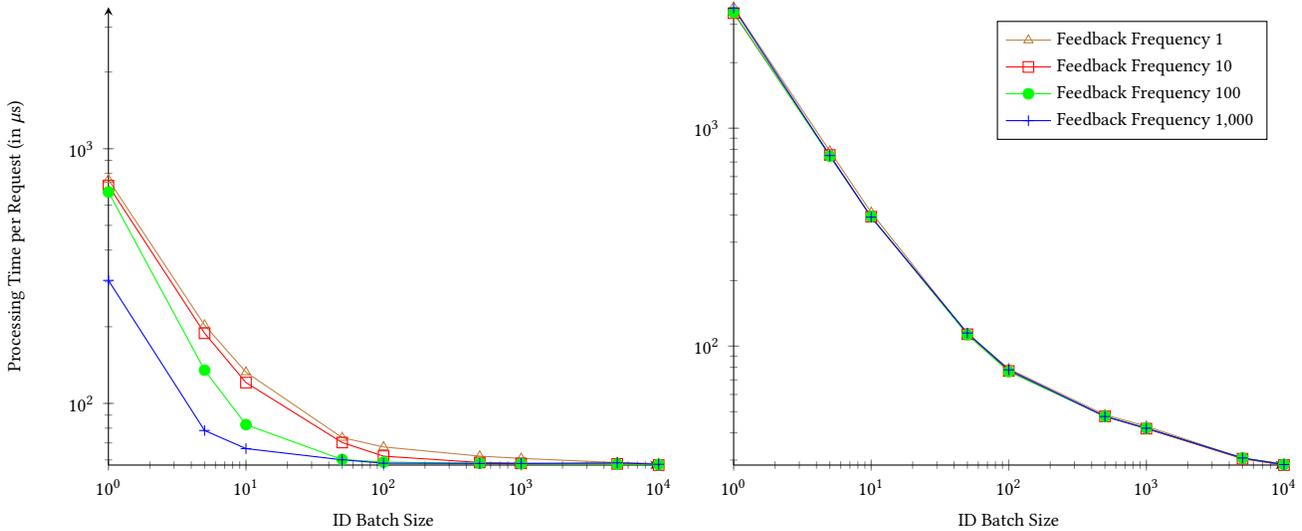

After measuring the processing time of the uninstrumented signing service, we evaluated the processing time when instrumenting the service with \name. 
Specifically, we measured the time required by the \pthread to process the request, cache the IDs produced by the \trampoline, and to store the encrypted \batches in the \queue. 
Additionally, we determined the time taken by the \vthread to read the encrypted \batches from the \queue, decrypt them, analyze the respective IDs, and store the result in the \rlog. 
Note that although the \vthread also spends time actively waiting for new \batches (\autoref{sec:implementation:execution}), we excluded this waiting time from our measurements in aspiration of an improved synchronization mechanism in the future.
 
\autoref{fig:eval:processing} depicts the time the \pthread requires to process a single request on the left and the time the \vthread requires to validate the control flow on the right. 
To
determine the impact of different \batch sizes and feedback frequencies, we again performed the measurements for different parameter combinations. 
Specifically, the X-axes show the different \batch sizes, while the Y-axes indicate the average processing time per request. 
Further, the graphs depict the different feedback frequencies. 
Similar to the evaluation of the \queue, 
the \ff has only limited impact on the processing time in comparison to the \batch size. 
Using a batch size of $1$, the \pthread requires between $755.84$ and $304.16$~$\mu$s to process a single request, depending on the \ff. 
Yet, the graph also shows that a higher \batch size can drastically reduce this processing time. 
For example, our default configuration of an \bsize of $10{,}000$ and a \ff of $10$, which we also used for the evaluation of the benchmark (\autoref{sec:evaluation:bperformance}), reduced the processing time to $57.28$~$\mu$s. 
This represents a $0.05$x overhead in comparison to the original response time of $54.79$~$\mu$s.
In \autoref{tab:evaluation:overhead}, we show the overhead for further configurations. 
While we recorded a maximum overhead of $12.80$x with both the \bsize and \ff set to $1$, we were able to reduce the overhead to as little as $0.05$x with different parameter combinations. 
This proves that while we recorded a considerable overhead under high CPU load, the impact on the \target remains limited for traditional applications such as our signing service.

\begin{table}[t]
	\centering
	\caption{
		Overheads for processing requests in the signing service for different \batch sizes and feedback frequencies.
	}
	\label{tab:evaluation:overhead}
	\begin{tabular}{lrrrr}
										& \multicolumn{4}{c}{\textbf{Feedback Frequency}} \\
										& \textbf{1}	& \textbf{10} 	& \textbf{100}    &	\textbf{1{,}000}\\
		\midrule
		\textbf{Batch Size 1}			& $12.80$x		& $12.06$x		& $11.34$x		  & $4.55$x \\
		\textbf{Batch Size 10}			& $1.42$x		& $1.20$x		& $0.50$x		  & $0.21$x \\
		\textbf{Batch Size 100}			& $0.23$x		& $0.13$x		& $0.07$x		  & $0.06$x \\
		\textbf{Batch Size 1,000}		& $0.11$x		& $0.06$x		& $0.06$x		  & $0.06$x \\
		\textbf{Batch Size 10,000}		& $0.06$x		& $0.05$x		& $0.05$x		  & $0.05$x \\
	\end{tabular}
\end{table}

Compared to the \pthread, the \vthread requires between $3{,}594.44$ and $3{,}554.69$~$\mu$s to process a single request with an \batch size of $1$. 
Again, the measurements show only a small difference between the different feedback frequencies. 
However, similar to the \pthread, we can drastically reduce the processing time per request by using a higher \bsize. 
For example, the default configuration of a \bsize of $10{,}000$ and a \ff of $10$ reduces the processing time to $28.46$~$\mu$s. %

In summary, our performance measurements show that both the \queue and the \analyzer are able to process the IDs faster than they are collected by the \trampoline. 
This eliminates the risk of creating a backlog of IDs to be analyzed when processing a large number of requests.  
Another conclusion from our measurements is that by caching IDs in the \prover and parallelizing the executing of the \target and the analysis, we can limit the overhead of the \gls{CFA} on the processing time of traditional applications to as little as $0.05$x.

\subsection{Security Evaluation}
\label{sec:evaluation:security}

To limit the performance impact of \name
our prototype caches IDs in the \prover.
In this section, we analyze the security impact of this caching mechanism and of reducing the feedback from the \verifier to the \prover.

Previous work~\cite{lee2017hacking,biondo2018guard} has shown that an attacker in control of a \gls{TEE} 
is able to 
execute arbitrary code within the \gls{TEE} using only a few \gls{ROP} gadgets. 
Executing such a gadget requires the attacker to take an invalid edge in the 
\target's 
\gls{CFG}. 
As we record the offset between the current instruction pointer and the jump destination (\autoref{sec:implementation:instrumentation}), jumping to a gadget will influence the recorded \callid. 
This allows us to detect the jump even if the gadget is located in uninstrumented code. 

To hide the attack, the attacker could delete or modify the ID created by the jump to the gadget while it is cached in the \prover.
\name detects such a modification by protecting cached IDs with a hash chain based on a secret provisioned to both \glspl{TEE} (\autoref{sec:implementation:instrumentation}). 
To recalculate the hash chain, the attacker has to be aware of this secret, which we delete immediately after calculating the first link. %
Hence, even an attacker able to execute arbitrary code cannot modify or delete cached IDs without being detected. 
Therefore, caching IDs in the \prover only influences the point in time at which recorded IDs are transferred to the \analyzer, but does not impair the security of \name. 

Another attack an adversary in control of the \prover could attempt would be to compromise the \verifier by sending it malicious \callids. 
To defend against such an attacker, we hardened the ID parser in the \verifier and reduced it to less than $30$ lines of code.

In addition to caching, our prototype can also reduce the feedback given to the \prover by adjusting the \ff.
To understand the impact of this parameter, we have to remember that we use a shared memory region created by the untrusted \gls{HA} to exchange \batches between the \prover and the \verifier. 
While the batches are encrypted and integrity protected, the \gls{HA} or a \rogueadmin could still block the transmission of all or specific \batches to the \verifier. 
The \ff allows us to detect such an interference. %
Specifically, the parameter states after how many \batches the \verifier acknowledges their receipt.   
This allows the \prover to detect when the \verifier does not receive sent batches. 
Similarly, the \ff also helps detecting a high privileged attacker pausing or stopping the \verifier. 
Specifically, if the \verifier does not acknowledge the receipt of previously sent \batches, the \prover can conclude that the collected IDs are not being verified and halt execution.

\section{Discussion}
\label{sec:discussion}

\name allows to detect control flow attacks in cloud environments. 
Depending on the particular requirements and possibilities, it is necessary to consider different aspects regarding its limitations and unavoidable attacks.

\paragraph{Missing Context-Sensitivity}
The current implementation of our prototype verifies the edges of the control flow, but does not consider its context. %
Hence, an attack mixing two different but valid control flows would remain undetected. 
For example, let us assume that the \target has two valid control flows: $A \rightarrow B \rightarrow C$ and $D \rightarrow B \rightarrow E$. 
If $B$ contains a vulnerability, the attacker 
can
execute the invalid control flow $A \rightarrow B \rightarrow E$. 
In this case, the analyzer would not 
be able to 
detect the attack, as it only considers the edges between two nodes in the \gls{CFG}, but does not analyze the previous control flow.  
Future work is required to determine how context-sensitive \gls{CFA} can be performed with an acceptable performance and storage overhead.

\paragraph{Integrity of the TEE}
Our design relies on both the \prover and the \verifier to provide confidentiality and integrity protection even in the presence of a \rogueadmin. 
However, previous research has shown that current implementations of \glspl{TEE} can be subject to attacks which violate these goals~\cite{wilke2020sevurity,nilsson2020survey,chen2021voltpillager,radev2020exploiting,wilke2021undeserved, morbitzer2021severity,buhren2021one}.
Such attacks will also have an effect on \name, as we would not be able to ensure the integrity of any of its components. 
Yet, the respective publications also discuss defense strategies, which we assume to be deployed when running \name. 

\paragraph{Launch-time Modification of \glspl{TEE}}
We rely on the untrusted \gls{HA} to launch both \glspl{TEE} 
(\autoref{sec:implementation:overview}).
This allows a malicious \gls{HA} to perform a Denial of Service attack by refusing to launch one or both \glspl{TEE}. 
Unfortunately, it is difficult to prevent such attacks in general. 
However, while not being able to prevent them, we can use the built-in static remote attestation for the respective \gls{TEE} to determine if both \glspl{TEE} have been launched~\cite{johnson2016intel, kaplan2016amd}.
Additionally, the static remote attestation allows us to verify whether the \glspl{TEE} have been launched without modifications.

\paragraph{Gadgets in Trusted Code}
Previous work showed that an attacker exploiting a vulnerability within an SGX enclave 
can
access a variety of gadgets in trusted code such as SGX libraries~\cite{lee2017hacking,biondo2018guard}.
Our prototype is able to detect such attacks by XORing the \callids of indirect branches with the offset between the current instruction pointer and the jump destination (\autoref{sec:implementation:instrumentation}). 
This safeguard allows us to detect an attacker overwriting the jump destination to point to trusted code, as this will influence the recorded \callid.

\paragraph{Manipulation of Cached IDs}
To reduce the overhead of transferring IDs from the \prover to the \verifier,
we cache collected IDs.
An attacker in control over the \target may attempt to modify or delete these IDs. 
Specifically, control over the \target would also imply access to the entire \prover, including the cached IDs.  
To protect the IDs, we add them to a hash chain (\autoref{sec:implementation:instrumentation}). 
After modification or deletion of cached IDs, the attacker will not be able to calculate a $hash_p'$ matching the new IDs. %
This is due to the fact that only the current value of $hash_p$ is accessible and the hash is not reversible. 
To nevertheless infer its previous values, a \rogueadmin could attempt to collect previously exchanged \batches containing earlier values of $hash_p$.  
Yet,
we encrypt the \batches before sending and use an irreversible \gls{KDF} to calculate a new encryption key for every exchange.
Hence, the attacker will not be able to decrypt previously sent \batches even when having access to the current encryption key. 
This ensures the integrity of IDs cached in the \batch even in the presence of a \rogueadmin 
who additionally gained control over the \prover.

\paragraph{Removing \Batches From the \Queue}
The \queue we use to exchange \batches between the \prover and the \verifier is located in shared memory. 
This enables a \rogueadmin to access batches stored in the \queue. 
To prevent inspection or modification of the \batches, we encrypt them with AES-GCM (\autoref{sec:implementation:overview}). 
Yet, an attacker may also attempt to remove an \batch from the \queue before it can be read by the \analyzer. 
We detect such an interference with the help of the counters in the \prover and \verifier used as IV for encryption and decryption of transmitted \batches. 
We update these counters after every data exchange. 
Therefore, if the \verifier does not receive an \batch sent by the \prover, the counters run out of sync. 
This means that if a batch is deleted from the \queue, the \verifier would attempt to decrypt the next batch with the wrong IV, detecting that a previous batch was not received.

\paragraph{Launching Multiple \gls{TEE} Instances.}
Another advantage of using counters as IV is that they allow us to detect a \rogueadmin launching two instances of the \verifier. 
In this scenario, the attacker could attempt to forward certain \batches indicating attacks to one instance of the \verifier and batches with regular IDs to another instance. 
Yet, such an attack would cause the counters in the \prover and the \verifier 
to run out of sync, allowing us to detect the attack. 
Additionally, we 
can also 
detect multiple instances of the \prover. 
Specifically, the \target in the \prover will be shipped without secrets such as the signature key required by our signing service. 
This is the normal procedure for enclave images, as the \gls{HA} has access to the image and 
could therefore
retrieve secrets shipped with the enclave. 
Instead, secrets are only provisioned after the enclave is successfully attested. 
Hence, when a malicious \gls{HA} launches multiple instances of the \prover, only one instance would be provisioned with the secret. 
This prevents the \gls{HA} from creating multiple functional instances of the \prover at the same time.

\section{Related Work}

In the past, a number of \gls{CFA} mechanisms have been proposed. 
While most of those mechanisms make use of \glspl{TEE} to protect the \gls{CFA}, we are not aware of any work that allows performing \gls{CFA} when the \target itself is also protected by a \gls{TEE}.

Abera et al.~\cite{abera2016cflat} first introduced the concept of \gls{CFA} in 2016. 
For each allowed control flow within the target program, they create a hash chain which contains the addresses of every executed basic block. 
While the hash chain allows the prover to store an entire control flow in a single hash, it also requires the verifier to store a hash for every possible flow. 
Further, they use addresses to identify basic blocks, which prevents the target from using \gls{ASLR}. 
In comparison, \name is compliant with \gls{ASLR} by using an incremental counter to uniquely identify the endpoints of edges in the \gls{CFG}.   
Also, instead of using hash chains to store valid control flows, we store the entire \gls{CFG}.
This facilitates identification of the basic block in which an attack has first taken effect.
Nevertheless, we also apply the concept of hash chains by using them to ensure the integrity of IDs cached in the \prover. 

Dessouky et al.~\cite{dessouky2017lofat} build on the work of Abera et al.~\cite{abera2016cflat}, but implement their design in hardware. 
This allows them to reduce the overall overhead and to attest the control flow without having to instrument the target program. 
Another hardware-based design by Zeitouni et al.~\cite{zeitouni2017atrium} additionally provides protection from physical attacks. 
Both 
realize the respective advantages by monitoring all executed instructions. 
Further work by Dessouky et al.~\cite{dessouky2018litehax} attempts to reduce the complexity when creating and verifying the \gls{CFG} by splitting it up into segments. 
In comparison, \name targets cloud environments, preventing us from performing hardware modifications. 
Yet, using the \glspl{TEE} offered by the cloud provider also allows us to defend against certain physical attacks such as DMA or cold boot attacks~\cite{boileau2006hit, halderman2008lest}. 

Abera et al.~\cite{abera2019diat} use \gls{CFA} to detect misbehavior of devices in distributed systems. 
In their work, they only attest critical code, reducing the performance overhead required to perform \gls{CFA}. 
Additionally, they reduce the storage overhead by using multiset hash functions. 
Conti et al.~\cite{conti2019radis} also apply \gls{CFA} to distributed systems, enabling them to detect services that perform non-intended operations. 
They achieve this goal by presenting a protocol which allows for the attestation of a control flow over multiple devices. 
Another design for distributed systems by Koutroumpouchos et al.~\cite{koutroumpouchos2019secure} proposes the application of Berkeley Packet Filters to extract the \gls{CFG} from devices.  
In contrast to these works, we target cloud environments, in which we only have to attest one target, yet this target is running on an untrusted system. 

Sun et al.~\cite{sun2020oat} use \gls{CFA} to verify the integrity of an operation executed on an embedded system. 
To reduce the overhead of the \gls{CFA}, they split the attestation report into a trace giving information about forward edges and a hash chain of return values ensuring backward edge protection. 
Compared to \name, this backward edge protection is not compatible with \gls{ASLR}.
Additionally, the authors only use \glspl{TEE} to protect recorded measurements, not to protect the \target itself. 

\gls{CFI} mechanisms take a slightly different approach and verify edges before their execution~\cite{burow19shining, burow17cfi, zieris18leak}. 
This requires meta data about the legal edges, which must be stored securely out of an attacker's reach. 
Existing \gls{CFI} mechanisms therefore aim to hide this information, though only with limited success~\cite{gawlik16enabling, goktacs16undermining, oikonomopoulos16poking}.
Considering our much stronger threat model of a \rogueadmin, securely hiding the meta data becomes even more difficult. 
We achieve this goal by storing the data for the verification in a separate \gls{TEE}, the \verifier, preventing access from both a \rogueclient and a \rogueadmin.

Toffalini et al.~\cite{toffalini2019scarr} also move the meta data required for the control flow verification into the verifier by introducing remote shadow stacks. 
Further, they reduce the complexity of the \gls{CFG} by splitting it into segments and reducing it to 
nodes critical for the target's execution. 
Similarly, Zhang et al.~\cite{zhang2021recfa} also move the control flow verification into the remote verifier.
Additionally, they simplify recorded control flows 
by skipping certain direct calls and folding control flow events such as loops and recursions. 
Compared to our design, the most important difference to these works is their weaker threat model. 
In detail, while we also assume a \rogueadmin, both Toffalini et al. and Zhang et al. use the kernel as trust anchor, making their design vulnerable to a high privileged adversary. 
With \name, we provide a method to securely verify the \target's control flow even under this strong threat model.

\section{Conclusion}

In this work, we presented \name, a design that allows to perform \gls{CFA} in \glspl{TEE}, created to be deployed in cloud environments.
\name verifies the control flow of a target hosted in a \gls{TEE} by making use of a second \gls{TEE}. 
While using \glspl{TEE} enables us to protect the target and the \gls{CFA} mechanism from a \rogueadmin, the separation into two \glspl{TEE} allows us to protect the \gls{CFA} mechanism from a potentially compromised target. 

We implemented our \name prototype using Intel SGX as \gls{TEE} and evaluated it in Microsoft Azure.
By deploying a caching mechanism protected by a hash chain, we significantly reduce the performance overhead of our prototype without sacrificing its security.  
A benchmark of the prototype shows that the additional \gls{CFA} may still require a considerable amount of time under high CPU load. 
Yet, we also show that the impact on traditional applications is limited.
Specifically, our evaluation of a signing service shows that by combining our caching mechanism with multithreading, we 
reduce
the overhead in the \target's response time to $0.05$x.

Another advantage of \name is that we only require the cloud provider to provision a \gls{TEE} and do not impose any further requirements. 
This makes our design suitable for the majority of popular cloud environments, enabling \sproviders to ensure the run-time integrity of their services even in such environments.

\begin{acks}
	This work was been funded by the Fraunhofer-Cluster of Excellence ``Cognitive Internet Technologies''. 
	We would like to thank Alina Weber-Hohengrund for integrating the mechanism for securely fetching the \rlog. 
\end{acks}

\bibliographystyle{ACM-Reference-Format}
\bibliography{biblio}

\end{document}